\documentclass[aps,pra,twocolumn,groupedaddress,nofootinbib,notitlepage,showpacs,floatfix,superscriptaddress]{revtex4-2}

\usepackage{amsmath}
\usepackage{amsfonts}
\usepackage{graphicx}
\usepackage{dcolumn}
\usepackage{relsize}
\usepackage{bm}
\usepackage{amsthm}
\usepackage{silence}
\usepackage{times}
\usepackage{color}
\usepackage{xcolor}
\usepackage[colorlinks=true,urlcolor=blue,citecolor=blue]{hyperref}
\usepackage{url}
\usepackage{adjustbox}
\usepackage{float}
\usepackage{soul}
\usepackage{graphicx}
\usepackage{amsmath}
\usepackage{appendix}
\usepackage[labelfont=bf,labelformat=empty,labelsep=space,position=auto]{subfig}
\usepackage{setspace}
\usepackage{ragged2e}
\usepackage{xcolor}
\usepackage{amssymb}
\usepackage{latexsym}
\usepackage{times}

\DeclareSubrefFormat{myparens}{#1(#2)}
\usepackage{hyperref}
\DeclareGraphicsExtensions{.png,.eps}
\usepackage{float}
\usepackage{ulem}
\usepackage{appendix}
\usepackage{etoolbox}

\usepackage{xcolor}
\usepackage[urlcolor=blue]{hyperref}      
\hypersetup{
	colorlinks = true,                    
	citecolor = {blue},
	linkcolor = {blue},
}

\begin{document}	
	
	\title{Nonreciprocal phonon blockade in spin quadratic optomechanical systems}
	
	\author{Yao Dong}
	\affiliation{School of Physics, Beihang University,100191,Beijing, China}
	
	\author{Guo-Feng Zhang}
	\email{gf1978zhang@buaa.edu.cn}
	\affiliation{School of Physics, Beihang University,100191,Beijing, China}

	\date{\today}
	
	\begin{abstract}
		We propose a scheme for achieving nonreciprocal phonon blockade in a quadratic optomechanical (QOM) system consisting of two spinning resonators near-field coupled to a nanomechanical oscillator. Due to the Sagnac–Fizeau effect, pump fields propagating in opposite directions experience distinct effective detunings, thereby leading to asymmetric intracavity intensities. Through the optical spring effect, this intensity imbalance gives rise to direction-dependent shifts in the effective mechanical frequency, providing the core mechanism for nonreciprocal phonon blockade. By judiciously setting parameters, single-phonon resonant excitation leads to conventional phonon blockade for one pump direction, whereas two-phonon resonance facilitates phonon-induced tunneling (PIT) for the other. The pronounced nonreciprocity is quantified by a contrast ratio in the phonon second-order correlation function exceeding 55 dB. To elucidate the nonreciprocal statistics, the phonon blockade is further analyzed in terms of interference between the coherent component and squeezed fluctuations. Incorporating thermal phonons, we reveal an extended nonreciprocal thermal effect, where increasing thermal noise degrades antibunching toward Poissonian statistics in one direction, yet reverses the statistics from bunching to antibunching in the opposite direction. Our work provides a pathway toward nonreciprocal phonon devices and directional phonon switches, with potential applications in chiral networks and phononic information processing.

	\end{abstract}
	
	\maketitle
	
	\section{INTRODUCTION}
	
	With the rapid development of quantum information technology and precision measurement, cavity optomechanics~\cite{RevModPhys.86.1391} has emerged as a core platform for exploring quantum effects at the macroscopic scale. Specifically, the radiation pressure of light provides a reliable way to create and manipulate phonons via the coherent coupling between optical and mechanical degrees of freedom. This optomechanical interaction enables the cooling of mechanical motion to its ground state~\cite{Chan2011,Clark2017,PhysRevX.8.041034,PhysRevLett.134.043601} and facilitates the generation of macroscopic quantum states, such as entanglement~\cite{Vitali2007,Ockeloen-Korppi2018,Riedinger2018,Li2018} and superpositions~\cite{Bild2023,PhysRevLett.117.143601,Liao2016,Weiss2024}, serving as a vital probe for the quantum-classical boundary. Furthermore, to fully exploit the potential of mechanical systems for quantum information processing, achieving control at the single-phonon level~\cite{OConnell2010,Hong2017,Enzian2021} is essential.
	
Phonon blockade (PB), a quantum effect analogous to the well-known Coulomb blockade~\cite{Kastner1992,PhysRevB.68.041311,McArdle2023} and photon blockade~\cite{Faraon2008,PhysRevLett.107.063601,PhysRevA.96.053810,PhysRevLett.134.183601}, manifests as the suppression of subsequent phonon excitations by the presence of a single phonon in the resonator. Operating in the PB regime, the mechanical resonator functions as a single-phonon source, playing a critical role in diverse quantum applications. For instance, single phonons can facilitate quantum frequency conversion and mediate microwave-to-optical quantum transduction, bridging superconducting quantum computing and distributed quantum networks~\cite{Andrews2014,Mirhosseini2020,Lauk2020}. Single-phonon sources~\cite{Nunnenkamp2011,Galland2014,Hong2017,Chu2018}, along with phonon beam splitters~\cite{Qiao2023,Weaver2017} and detectors~\cite{Cohen2015,Sletten2019,Lecocq2015}, are fundamental enabling components in linear mechanical quantum computing~\cite{Qiao2023}. Moreover, the single-phonon state has the potential to enhance sensitivity in quantum metrology for force and displacement~\cite{Degen2017,Wolf2019}. Based on the underlying mechanism, PB can be categorized as conventional (driven by nonlinearity) or unconventional (induced by destructive quantum interference), mirroring the established classification of photon blockade~\cite{PhysRevA.102.033713,PhysRevLett.121.043601,Casalengua2020,PhysRevLett.134.183601}. Despite the greater experimental challenges associated with phonon blockade compared to photon blockade, extensive research has led to remarkable advances. By coupling mechanical resonators to hybrid platforms—including superconducting qubits~\cite{Didier2011,PhysRevA.82.032101,PhysRevA.93.013808,PhysRevA.94.063853,Chu2017,PhysRevApplied.17.054004}, intrinsic two-level defects~\cite{PhysRevLett.110.193602}, and nitrogen-vacancy centers~\cite{PhysRevLett.110.156402,PhysRevA.100.063840,Cai2018,PhysRevA.111.052620}—or exploiting quadratic optomechanical interactions~\cite{PhysRevA.96.013861,PhysRevA.98.013821,PhysRevA.98.023819,PhysRevA.99.013804,Wei2023}, the realization of robust single-phonon sources is becoming increasingly feasible. In particular, quadratic optomechanics is a promising platform for PB, since the typically weak single-photon quadratic coupling can be coherently enhanced to yield an effective strong nonlinearity.
	
In parallel with these advances, breaking time-reversal symmetry to realize nonreciprocal, direction-dependent transport has garnered significant attention~\cite{Jalas2013,Estep2014,PhysRevLett.130.203801,PhysRevA.111.023510,Verhagen2017}. Nonreciprocal devices, which effectively block back-reflected signals, are indispensable for one-way optical communications and noise isolation~\cite{Jalas2013,Gisin2007,Bi2011,Yu2009}. To circumvent the limitations of conventional magnet-based approaches, which are typically bulky and incompatible with device miniaturization~\cite{PhysRevLett.109.033901,Hua2016,Verhagen2017,Sounas2017}, alternative strategies exploiting quantum nonlinearity~\cite{DelBino2018,PhysRevLett.121.123601,PhysRevLett.123.233604,PhysRevApplied.16.014046}, chiral optics~\cite{PhysRevLett.110.213604,Sollner2015,Qie2023}, parity-time ($\mathcal{PT}$)-symmetric~\cite{PhysRevLett.103.093902,Peng2014,Chang2014} and optomechanics~\cite{Ruesink2016,Bernier2017,Fang2017,PhysRevLett.125.023603,PhysRevLett.130.013601} have recently been experimentally demonstrated. Subsequently, extending beyond classical energy transmission, the concept of nonreciprocity has been expanded to the regime of quantum statistics, specifically nonreciprocal photon blockade~\cite{PhysRevLett.121.153601}. In this scenario, the system behaves as a single-photon source for one input direction, while functioning as a conventional or even super-Poissonian source for the opposite direction. To date, most schemes for realizing nonreciprocal photon blockade have been implemented in spinning~\cite{PhysRevLett.121.153601,Li2019,PhysRevA.100.053832,PhysRevA.104.033707,PhysRevA.108.043723,Zuo2024} or static~\cite{Xu:20,PhysRevA.110.023723,PhysRevA.111.033715,PhysRevA.106.053707} whispering-gallery-mode (WGM) resonators. The former relies on the Fizeau shift between clockwise (CW) and counterclockwise (CCW) modes, which arises from the optical Sagnac effect. This frequency difference induces an asymmetric eigenenergy spectrum for conventional blockade or distinct quantum interference conditions for unconventional blockade. In the latter case, blockade arises exclusively in the target direction due to directional interactions, including optomechanical nonlinearity~\cite{Xu:20}, parametric amplification~\cite{PhysRevA.110.023723}, chiral cavity–atom interaction~\cite{PhysRevA.111.033715}, and optomagnonic coupling~\cite{PhysRevA.106.053707}. Although nonreciprocal photon blockade and direction-independent phonon blockade have been extensively studied,  the manipulation of direction-dependent phonon statistics via optical driving, particularly in quadratic optomechanical systems involving high-order interactions, has not been explored.  

 In this work, we propose a scheme to achieve nonreciprocal PB in QOM systems with spinning WGM resonators, where the required quadratic coupling emerges from the avoided crossing of optical supermodes. Coherent driving of the resonators by a pump optical field induces an effective nonlinear coupling between photons and phonons, resulting in an anharmonic energy spectrum, which is a prerequisite for PB in our proposed system. For pump fields propagating along or against the rotation direction, the Sagnac-Fizeau effect induces opposite frequency shifts, leading to direction-dependent pump detunings. This results in asymmetric intracavity intensities and is ultimately imprinted onto the effective mechanical frequency via the optical spring effect. Specifically, when the single-phonon energy in one direction matches half the energy of a two-phonon eigenstate in the opposite direction, the applied weak mechanical drive resonantly excites a single-phonon transition in the former case, whereas it induces a two-phonon resonance in the latter. Consequently, nonreciprocal phonon statistics emerge in this spinning QOM system, manifesting as PB for pump light input from Port 1 and PIT for input from Port 2. By simultaneously satisfying this target nonreciprocal PB criterion and the photon-two-phonon resonance,  we numerically determine the optimal resonator spinning angular velocity and pump power for the specified QOM system. We characterize the phonon statistics via the equal-time second-order correlation function. The results obtained from the analytical wave-function approximation shows good consistency with numerical master-equation simulations. Furthermore, drawing inspiration from advanced studies on photon statistics, we provide an interpretation of the nonreciprocal PB in terms of quantum interference between the coherent component and squeezed fluctuations. Our work provides a route to realize nonreciprocal single-phonon devices, which can be flexibly reconfigured by switching the pump port and have promising applications in hybrid photon-phonon quantum networks~\cite{Dong2015} and chiral quantum technologies~\cite{Lodahl2017}.

	\section{THEORETICAL FRAMEWORK AND SYSTEM HAMILTONIAN}\label{section2}

	\begin{figure}[tbp]
		\centering
		\subfloat[]{\includegraphics[width=1\linewidth]{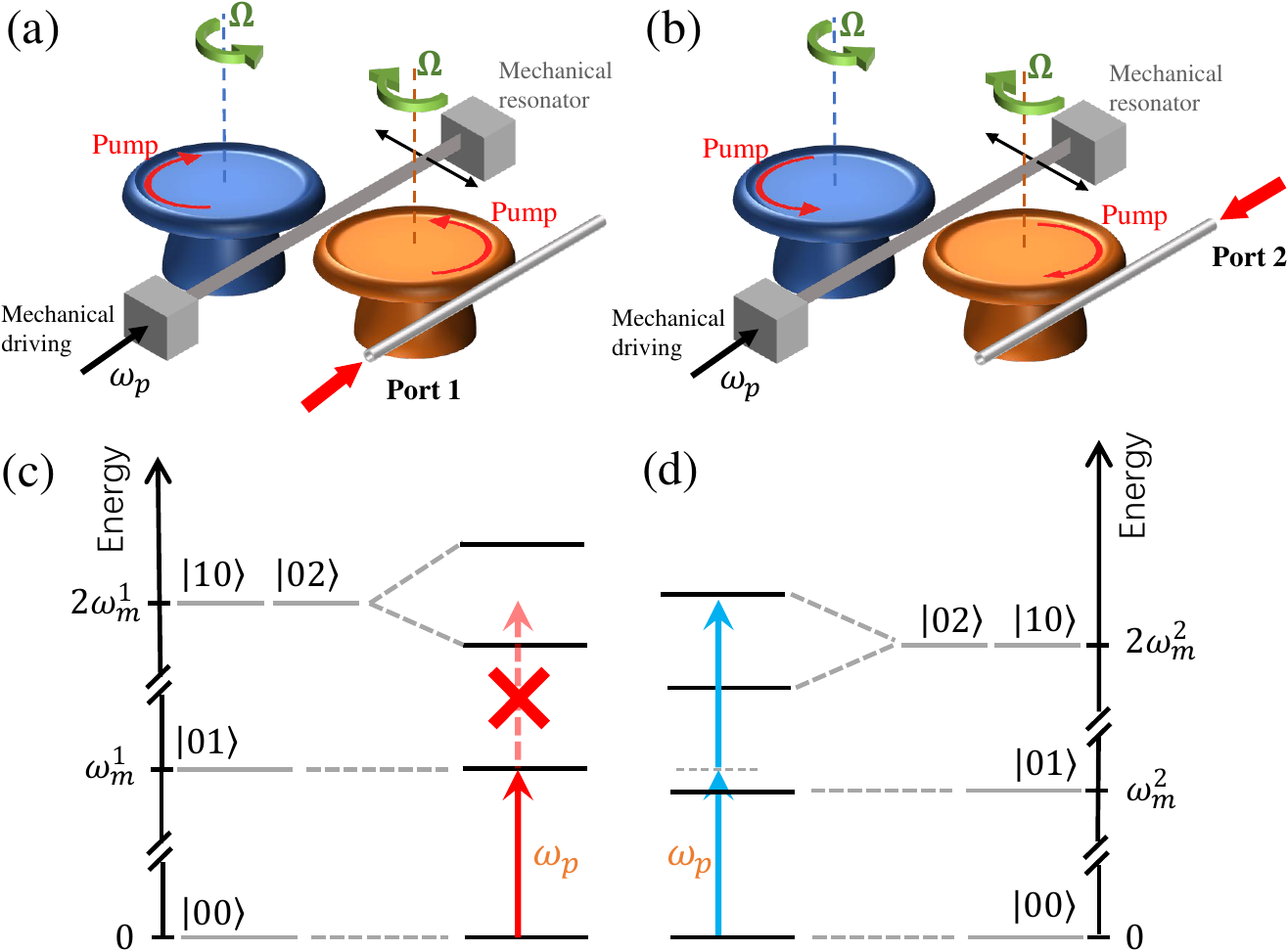}\label{schematic}}\\ \vspace{-5mm}%

		\caption{\justifying{(a), (b) Schematic of a spinning-resonator system for generating QOM coupling and nonreciprocal PB. Two WGM resonators rotate with the same angular velocity in opposite directions, coupled to a central mechanical oscillator. A coherent pump field is input from port 1 [in (a)] or port 2 [in (b)], propagating against or along the rotation direction of the cavities. To excite phonons in this nonreciprocal phonon source, a weak driving field is applied to the mechanical mode. (c), (d) Eigenenergy-level diagrams for the configurations in (a) and (b). $|n,m\rangle$ represents a state with $n$ photons and $m$ phonons. The effective optomechanical interaction lifts the degeneracy of the eigenstates in the two-phonon subspace. In (c), PB arises from the energy-level anharmonicity under single-phonon resonance (red arrows), while in (d), PIT is induced by two-phonon resonance (blue arrows).}}\label{Fig.1}
	\end{figure}
We consider a compound optomechanical system, as shown in Fig.~\ref{Fig.1}, which consists of two spinning optical resonators coupled to a mechanical resonator via the evanescent near field~\cite{Anetsberger2009,Schilling2016,Xu:20}. Specifically, the two optical resonators rotate with angular velocities $\Omega$ in clockwise and counterclockwise directions, respectively. According to the Sagnac effect,  the resonance frequency of light circulating in each spinning resonator undergoes a Sagnac-Fizeau shift, given by $\omega_c\to\omega_c+\Delta_F$~\cite{Malykin2000}, where
\begin{align}\label{Delta_F}
	\Delta_F = \pm \dfrac{nr\Omega\omega_c}{c}\left(1-\dfrac{1}{n^2}-\dfrac{\lambda}{n}\dfrac{dn}{d\lambda}\right).
\end{align}
Here, $\omega_c$ is the resonance frequency of the stationary  resonator, $c$ and $\lambda$ are the speed and wavelength of light in vacuum, $r$ is the  resonator radius, and $n$ is its refractive index. The last term in Eq.~(\ref{Delta_F}), which characterizes the relativistic origin of the Sagnac effect, can be neglected as it is relatively small in typical experimental materials~\cite{Maayani2018}. A positive $\Delta_F$ denotes counter-rotating light propagation relative to the resonator spin (Fig.~\ref{Fig.1}(a)), whereas negative value ($\Delta_F < 0$) indicates co-rotation (Fig.~\ref{Fig.1}(b)). The system is described by the Hamiltonian 
\begin{align}\label{Hamiltonian}
	H=&\sum_{i=1,2}\left [ \omega_c\pm |\Delta_F|+(-1)^i g_0q\right ] a_i^\dagger a_i\notag\\&+J\left (a_1^\dagger a_2+a_1a^\dagger_2\right)+\dfrac{1}{2}\omega_m(p^2+q^2),
	\end{align}
 where $a_i$ (or $a_i^\dagger$) is the annihilation (or creation) operator for the $i$-th optical mode, $q$ and $p$ represent  the dimensionless position and momentum operators of the mechanical mode, and $J$ denotes the tunneling coupling strength. Since the gradient forces exerted by the two WGM microcavities on the central mechanical resonator act in opposite directions, the linear single-photon optomechanical coupling rates take the same magnitude $g_0$ but opposite signs. This form of interaction Hamiltonian has also been investigated in membrane-in-the-middle cavity optomechanical systems~\cite{PhysRevA.77.033819,PhysRevA.81.011801,PhysRevLett.103.100402}, photonic-crystal setups~\cite{PhysRevX.5.041024}, and WGM microcavities with a membrane placed above~\cite{PhysRevA.85.053832}.
 
 Under the assumption that $J \gg \omega_m$ and the weak optomechanical coupling limit $J \gg g_0 q$, the mechanical displacement $q$ behaves as a quasistatic variable~\cite{PhysRevLett.103.100402,PhysRevX.5.041024,PhysRevLett.109.063601}.By diagonalizing the Hamiltonian and applying a Taylor expansion, the linear coupling terms explicitly cancel out due to the supermode symmetry preserved by the oppositely rotating configuration of the resonators(the physical consequences of breaking this symmetry are discussed in Sec.~\ref{section4}), yielding the effective quadratic coupling (see Appendix~\ref{Appendix} for detailed derivation). The effective Hamiltonian is thus written as:
\begin{align}\label{Eq3}
	H=&\left ( \omega_c\pm|\Delta_F|+J+G_0q^2 \right)a_+^\dagger a_+\notag\\
	&+\left ( \omega_c\pm|\Delta_F|-J-G_0q^2 \right)a_-^\dagger a_-\notag\\
	&+\dfrac{1}{2}\omega_m(p^2+q^2),
\end{align}
 where $a_\pm\approx\left(a_1\pm a_2\right)/\sqrt{2}$ are the normal modes and $G_0=g_0^2/(2J)$ is the  quadratic optomechanical coupling coefficient. The tunneling coupling $J$ induces an anticrossing of the eigenfrequencies $\omega_\pm(q)\approx\omega_c\pm|\Delta_F|\pm(J+G_0q^2) $, where the $\pm$ signs apply exclusively to the last term. The experiment in Ref.~\cite{PhysRevX.5.041024} demonstrates that, when the optical mode splitting $2J$ significantly exceeds the cavity decay rate $\kappa$, the resonance dips in the probe laser transmission spectrum appear at the supermode resonance frequencies $\omega_\pm(0)$ instead of the bare cavity frequency. Consequently, the normal modes couple to pump fields injected through either port 1 or port 2. By tuning the strong pump field frequency $\omega_L$ to selectively excite the odd supermode  $a_-$ (labeled $A$ hereafter) with frequency $\omega_a=\omega_-(0)$,  the effective optomechanical coupling between mode $A$  and the mechanical resonator is parametrically enhanced. The system Hamiltonian in the rotating frame defined by the pump frequency $\omega_L$ is given by ($\hbar=1$)
 \begin{align}
 	H_{\text{om}}^{\text{pum-i}}=&\Delta_a^iA^\dagger A+\dfrac{1}{2}\omega_m(p^2+q^2)-G_0A^\dagger Aq^2\notag\\
 	&+E\left(A+A^\dagger\right),
 	\end{align}
where the superscript $i$ denotes that the strong pump is injected into port $i$, $\Delta_a^i=\omega_c-J-(-1)^i|\Delta_F|-\omega_L$ is the frequency detuning, and $E=\sqrt{\kappa P_{\text{in}}/(\hbar\omega_L)}$ is the driving amplitude, with  $\kappa$ being the cavity decay rate and $P_{\text{in}}$ the input power. In the strong driving regime, we linearize the system by decomposing each operator into its steady-state mean value and a small quantum fluctuation operator:  $A\to\alpha_i+A$, $q\to q_s+q$, and $p\to p_s+p$.  By solving the quantum Langevin equations following standard mean-field methods in quadratic optomechanics~\cite{PhysRevA.93.063860,Xu:20,PhysRevA.98.013821}, the steady-state values are given by:
\begin{align}\label{alphai}
	\alpha_i=\dfrac{-2iE}{\kappa+2i\Delta_a^i},
\end{align}
 and 
 \begin{align}
 	p_s = q_s =0.
 \end{align}
 We now consider the case where the mechanical mode is also driven by a weak field with amplitude $\varepsilon$ and frequency $\omega_p$. In the basis of the phonon creation and annihilation operators $b^\dagger $ and $b$, related to the mechanical quadrature by $q=(b+b^\dagger)/\sqrt{2}$ and $p = i(b^\dagger-b)/\sqrt{2}$, the effective Hamiltonian governing the quantum fluctuations for a pump incident from port-i takes the form:
\begin{align}
	H_\mathrm{eff}^i &= \Delta_a^i
	A^\dagger A+\left( \omega_m-G_0|\alpha_i|^2\right )b^\dagger b-\dfrac{G_0}{2}|\alpha_i|^2\left (  b^2+b^{\dagger 2}\right )\notag\\
	&\quad - \dfrac{G_0}{2}\left ( \alpha_i A^\dagger +\alpha_i^* A \right )(b+b^\dagger)^2\notag\\
	&\quad -\dfrac{G_0}{2}A^\dagger A\left ( b+b^\dagger \right )^2 +\varepsilon \left( b^\dagger e^{-i \omega_p t}+b e^{i\omega_p t}\right).  
	\end{align}
 The second term in the Hamiltonian describes the mechanical oscillator with an effective frequency
 \begin{align} \label{omegai}
 	\omega_m^i=\omega_m-G_0|\alpha_i|^2. 
 \end{align}
 	This frequency shift is a manifestation of the optical spring effect, arising from the intracavity field. For  a sufficiently strong optical drive $\left(|\alpha_i|^2\gg\langle A^\dagger A\rangle \right)$, the small second-order term  $G_0 A^\dagger A \left(b+b^\dagger\right)^2$, which lacks coherent enhancement from the classical field, can be safely neglected. Furthermore, in the case of $\Delta_a^i\sim 2\omega_m^i$ and  $\omega_m^i\gg G_0|\alpha_i|$, we can apply the rotating-wave approximation (RWA). This allows us to neglect the rapidly oscillating terms with frequencies of $2\omega_m^i$ and $4\omega_m^i$~\cite{PhysRevA.96.013861,PhysRevA.98.013821,PhysRevA.98.023819,PhysRevA.99.013804,Wei2023}. In the rotating reference frame defined by the unitary transformation $U(t) = \mathrm{exp}\left(i2\omega_p A^\dagger A+i\omega_p b^\dagger bt\right)$, the simplified effective Hamiltonian takes the form as
\begin{align}\label{sim_H}
	H^i_\mathrm{eff} &= 2\Delta_p^i A^\dagger A+\Delta_p^i b^\dagger b+G_iA^\dagger b^2+G_i^*Ab^{\dagger 2}\notag\\
	&\quad +\varepsilon\left(b+b^\dagger\right),
\end{align}
 where the detuning $\Delta_p^i=\omega_m^i-\omega_p$ satisfies the condition $|\Delta_p^i|\ll\omega_m^i$, and $G_i = |\alpha_i| G_0/2 $ is the effective second-order nonlinear coupling strength which we take to be real without loss of generality. 
 
 To incorporate the effects of dissipation, the dynamics of the system is governed by the Lindblad master equation for the density matrix $\rho$,
 \begin{align} \label{master}
 	\partial _t\rho&=i[\rho,H_{\mathrm{eff}}^i] +\kappa \mathcal{L} \left [ A \right ] \rho\notag \\
 	&\quad +\gamma\left ( n_\mathrm{th}+1  \right ) \mathcal{L} \left [ b \right ] \rho + \gamma n_\mathrm{th}\mathcal{L}\left [ b^\dagger \right ] \rho,
 \end{align} 
 where $\mathcal{L}[o]\rho=o\rho o^\dagger-\frac{1}{2}\left\{o^\dagger o,\rho\right\} $ is the Lindblad superoperator for an operator $o$, $\gamma$ is the mechanical dissipation rate. The mechanical resonator is coupled to a thermal environment at temperature $T$, which corresponds to a mean thermal phonon number  $n_{\mathrm{th}}=1/\left [ \mathrm{exp } \left ( \hbar\omega_m/k_B T\right ) -1\right ]$. The statistical properties of the phonon are characterized by the steady-state second-order correlation function as
 \begin{align}
 	g_b^{(2)}(0)=\lim_{t \to \infty} \dfrac{\langle b(t)^\dagger b(t)^\dagger b(t) b(t)\rangle  }{\langle b(t)^\dagger b(t)\rangle^2}=\dfrac{\mathrm{Tr }\left ( \rho_{ss} b^\dagger b^\dagger bb\right )}{\left [ \mathrm{Tr}\left ( \rho_{ss}b^\dagger b \right )   \right ]^2 },
 \end{align}
where the steady-state density matrix $\rho_{ss}$ is the long-time solution to the master equation presented in Eq.~(\ref{master}). In direct analogy to the statistics of an optical field, a value of $g_b^{(2)}(0)>1$  corresponds to super-Poissonian statistics, indicating phonon bunching. Conversely, the condition $g_b^{(2)}(0)<1$  implies sub-Poissonian statistics and serves as direct evidence of the nonclassical effect of phonon antibunching. Importantly, $g_b^{(2)}(0)\to0$ is a signature of strong phonon blockade, where only a single phonon can be excited in the mechanical resonator. Experimentally, the non-classical nature of the single phonon state can be verified by measuring the second-order correlation function using a Hanbury-Brown-Twiss (HBT) type setup~\cite{Hong2017}.
    \section{NONRECIPROCAL PHONON STATISTICS}\label{section3}
    \subsection{Mechanism and optimal conditions}
    In this section, we investigate the physical mechanism and the required conditions for realizing nonreciprocal phonon blockade, where the statistical properties of phonons depend on the direction of the incident optical pump. Firstly, the Sagnac-Fizeau effect induces a direction-dependent frequency shift, leading to different detunings of the pump field when injected through port 1 versus port 2: $\Delta_a^1 = \Delta_c + |\Delta_F|$ and $\Delta_a^2 = \Delta_c - |\Delta_F|$, where $\Delta_c = \omega_c - J - \omega_L$. This asymmetry in detuning, with a total difference of $2|\Delta_F|$, directly leads to different steady-state amplitudes of the intracavity field. Under the assumption that $\Delta_a^i \gg \kappa$, these amplitudes can be approximated from Eq.~(\ref{alphai}) as $\alpha_1 \sim -E/(\Delta_c+|\Delta_F|)$ and $\alpha_2 \sim -E/(\Delta_c-|\Delta_F|)$. By substituting asymmetric amplitudes into Eq.~(\ref{omegai}), this asymmetry directly translates into a nonreciprocal optical spring effect, explicitly yielding the direction-dependent effective mechanical frequencies
    \begin{align}\label{eff_mech1_fre}
    	\omega_m^1 = \omega_m-G_0 \dfrac{E^2}{\left ( \Delta_c+|\Delta_F| \right )^2 }
    \end{align}
    and 
    \begin{align}\label{eff_mech2_fre}
    	\omega_m^2 = \omega_m-G_0 \dfrac{E^2}{\left ( \Delta_c-|\Delta_F| \right )^2 },
    \end{align}
    as plotted in Figs.~\ref{Fig.1}(c) and~\ref{Fig.1}(d). The anharmonic energy spectrum of the effective Hamiltonian in Eq.~(\ref{sim_H}) is also illustrated in Figs.~\ref{Fig.1}(c) and~\ref{Fig.1}(d). The state $|n, m\rangle$ in the bare Fock basis corresponds to $n$ photons in optical mode $A$ and $m$ phonons in mechanical mode $b$. By diagonalizing the Hamiltonian (Eq.~(\ref{sim_H})) in the absence of the weak classical drive, we obtain the dressed eigenstates of the system,  which arise from the resonant interaction between the  states $|n,m\rangle$ and $|n-1, m+2\rangle$. Within the low-excitation subspace (up to two phonons), the relevant eigenstates are given by
    \begin{align} |0_0\rangle = |0,0\rangle, \ |1_0\rangle = |0,1\rangle, \ |2_\pm\rangle = \frac{1}{\sqrt{2}}(|1,0\rangle \pm |0,2\rangle). \end{align}
   The optomechanical interaction lifts the degeneracy of the two-excitation subspace, resulting in an energy splitting for the dressed states $|2_\pm\rangle$. Their corresponding eigenvalues are shifted to $E_{2\pm} \approx 2\omega_m^i \pm \sqrt{2}G_i$.
   
    We first consider the scenario where the pump is injected through port 1, with the mechanical drive tuned to be resonant with the single-phonon transition, i.e., $\omega_p = \omega_m^1$. This condition allows for the resonant excitation of the system from the ground state $|0_0\rangle$ to the first excited state $|1_0\rangle$. However, due to the anharmonicity of the energy ladder, the subsequent transition $|1_0\rangle \to |2_\pm\rangle$ is strongly detuned from the drive frequency,  as depicted by the red arrows in Fig.~\ref{Fig.1}(c). The excitation of the second phonon is therefore suppressed, resulting in  the conventional phonon blockade effect. When pumping through port 2, the nonreciprocal shift in the effective mechanical allows the same drive ($\omega_p = \omega_m^1$) to resonantly excite the two-phonon transition $|0_0\rangle \to |2_+\rangle$, as shown with the blue arrows in Fig.~\ref{Fig.1}(d). In stark contrast to the single-phonon blockade observed for the opposite pump direction, this process results in phonon bunching. Then the optimal resonator spinning angular velocity and pump strength are determined by simultaneously satisfying the two-phonon resonance condition ($\Delta_a^i=2\omega_m^i$) and connecting the nonreciprocal blockade and bunching mechanisms. This leads to the following set of equations: 
    \begin{align}\label{opt_condition}
    	&\dfrac{\Delta_c+\Delta_F}{2}=\omega_m-G_0 \dfrac{E^2}{\left ( \Delta_c+|\Delta_F| \right )^2 },\notag\\
    	&\dfrac{\Delta_c-\Delta_F}{2}=\omega_m-G_0 \dfrac{E^2}{\left ( \Delta_c-|\Delta_F| \right )^2 },\notag\\
    	&2\omega_m^1=2\omega_m^2+\sqrt{2}G_2.
    \end{align}
    For two identical resonators, the experimentally accessible parameters are chosen as follows~\cite{PhysRevLett.121.153601,PhysRevA.71.013817,PhysRevLett.116.133902,Pavlov2017}: $n=1.4$, $r=30\,\mu\text{m}$, $\omega_c=2\pi\times200\,\text{THz}$, and $Q=5\times 10^9$ (typically in the range $10^9\!-\!10^{12}$~\cite{PhysRevLett.116.133902,Pavlov2017}). The mechanical oscillator under consideration is a doubly clamped nanobeam fabricated from a high-stress $\mathrm{Si_3N_4}$ thin film~\cite{Verbridge2006,Anetsberger2009,Schilling2016}. With dimensions of $7\,\mu\mathrm{m} \times 500\,\mathrm{nm} \times 100\,\mathrm{nm}$ ($l \times w \times h$), the resonator possesses a fundamental frequency $\omega_m \sim 2\pi \times 40\,\mathrm{MHz}$ and a room-temperature quality factor $Q_m$ exceeding $10^5$~\cite{Verbridge2006}. Alternatively, phononic crystal nanobeams provide an attractive platform~\cite{Zhang2017,Burek2016}, offering ready experimental access to this frequency regime. By precisely controlling the separation between the two WGM resonators~\cite{Peng2014,Chang2014}, the inter-cavity optical coupling strength is set to $J/2\pi = 1000 \text{ MHz}$. This coupling strength satisfies the condition $J \gg \omega_m, \kappa$, ensuring that the system operates in the resolved supermode regime. This regime is essential for validating the effective quadratic optomechanical interaction and allows for selectively driving the target supermodes. We set $G_0 \equiv  \frac{g_0^2}{2J} = \kappa/20$, corresponding to a vacuum single-photon optomechanical coupling rate of $g_0/2\pi =  2 \text{ MHz}$. Although the coupling strength reported in near-field WGM experiments a decade ago was typically in the hundred-kHz regime~\cite{Schilling2016}, the desired $g_0$ is foreseeable thanks to recent advances in quantum enhancement schemes and modern nanofabrication capabilities. On one hand, various schemes including periodic arrays of mechanical oscillators~\cite{PhysRevLett.109.223601}, optical squeezing~\cite{PhysRevLett.114.093602}, and PT-symmetry~\cite{PhysRevLett.117.110802} have been proposed to enhance $g_0$. Utilizing these proposals, the optomechanical coupling $g_0$ can be improved by several orders of magnitude. On the other hand, beyond WGM-based systems, phononic crystal nanobeams provide a powerful route to enhance the vacuum optomechanical coupling by strongly reducing the effective mechanical mass of localized defect modes. While bare coupling strengths exceeding $10 \text{ MHz}$ have been demonstrated in standing-wave phononic crystal cavities with sliced nanobeam designs~\cite{Leijssen2015}, incorporating such structures into a high-speed spinning geometry presents significantly greater experimental challenges compared to traveling-wave WGM resonators. Therefore, in this section, we utilize $g_0/2\pi = 2 \text{ MHz}$ as an idealized limit to theoretically demonstrate the maximum potential and optimized quantum statistics of the nonreciprocal mechanism. We will subsequently assess the experimental viability of our scheme under realistic, unenhanced  WGM coupling parameters (i.e., $g_0/2\pi = 200 \text{ kHz}$) in Sec.~\ref{sec3_c}. By solving Eq.~(\ref{opt_condition}) with the system parameters discussed above, we determine the optimal spinning angular velocity $\Omega = 4.2 \text{ kHz}$ and pump power $P_{\text{in}} = 0.395 \text{ mW}$.
    
   To analytically solve for the second-order correlation function, we employ an effective non-Hermitian Hamiltonian that accounts for dissipation:
   \begin{align}
   	{H_{\mathrm{eff} }^i}' = H_{\mathrm{eff}}^i - i\frac{\kappa}{2}\hat{A}^\dagger \hat{A} - i\frac{\gamma}{2}\hat{b}^\dagger \hat{b}.
   \end{align}
   For a weak mechanical drive, the system remains in the low-excitation subspace, allowing us to truncate the Hilbert space and expand the system state as $|\psi(t)\rangle = \sum_{n,m} C_{nm}(t)|n,m\rangle$, where $C_{nm}$ denote the probability amplitudes, and the sum is over the basis states $\{|0,0\rangle, |0,1\rangle, |1,0\rangle, |0,2\rangle\}$. The dynamics of the probability amplitudes are thus governed by the following system of linear differential equations:
    \begin{align}
    	&i \dot{C}_{00}(t)= \varepsilon C_{01}(t),\notag \\
    	&i \dot{C}_{01}(t)= \left(\Delta_p^i -\dfrac{i\gamma}{2}\right)C_{01}(t)+\varepsilon C_{00}(t)+\sqrt{2}\varepsilon C_{02}(t),\notag\\
    	&i \dot{C}_{02}(t)= \left(2\Delta_p^i - i\gamma \right)C_{02}(t)+\sqrt{2}G_iC_{10}(t)+\sqrt{2}\varepsilon C_{01}(t),\notag\\
    	&i \dot{C}_{10}(t)= \left(2\Delta_p^i - \dfrac{i\kappa}{2} \right)C_{10}(t)+\sqrt{2}G_iC_{02}(t).
    	\end{align}
  In the limit $\varepsilon/\kappa\to 0$, the steady-state solutions can be obtained by setting all time derivatives in the equations of motion to zero, which leads to 
  {\allowdisplaybreaks
  \begin{align}
  	&C_{01}(\infty )=\dfrac{-\varepsilon }{\Delta_p^i-i\gamma/2} ,\notag\\
  	&C_{02}(\infty )=\dfrac{-2\sqrt{2} \varepsilon^2\left ( 4\Delta^i_p- i\kappa\right )  }{\left ( 2\Delta_p^i-i\gamma \right )\left [ 4G_i^2+\left ( \gamma+2i\Delta_p^i  \right ) \left ( \kappa+4i\Delta_p^i \right )  \right ]  } ,\notag\\
  	&C_{10}(\infty )=\dfrac{ 8 G_i\varepsilon ^2}{\left ( 2\Delta_p^i-i\gamma \right )\left [ 4G_i^2+\left ( \gamma+2i\Delta_p^i  \right ) \left ( \kappa+4i\Delta_p^i \right )  \right ]  } .
  \end{align}}
  Then, the second-order correlation function of the phonon mode, when the pump is incident on port i in the weak-driving regime, can be analytically expressed as
  \begin{align}
  	g_i^{(2)}(0)&=\dfrac{2|C_{02}|^2}{\left ( |C_{01}|^2 +|C_{02}|^2\right )^2 }\approx \dfrac{2|C_{02}|^2}{|C_{01}|^4}\notag \\
  	&\quad \approx \dfrac{|\left ( 4i\Delta_p^i+\kappa \right )\left (2i\Delta_p^i+\gamma  \right )  |^2}{|4G_i^2+\left ( 4i\Delta_p^i+\kappa \right )\left (2i\Delta_p^i+\gamma  \right )|^2}.
  \end{align}
 To quantify the degree of nonreciprocity between the phonon blockade and bunching effects observed for the two input ports, we define the contrast ratio $I_b$ in decibels (dB) as
   \begin{align} 
   	I_b = -10\log_{10}\left[ \frac{g_1^{(2)}(0)}{g_2^{(2)}(0)} \right]. 
   \end{align}
   Based on the preceding theoretical analysis, the nonreciprocal isolation $I_b$ is maximized under the conditions $\Delta_p^1 = 0$ and $\Delta_p^2 = -\sqrt{2}G_2/2$.

  \subsection{Nonreciprocal statistics and quantum interference} 
\begin{figure}[tbp]
   	\centering
   	\subfloat[]{\includegraphics[width=0.47\linewidth]{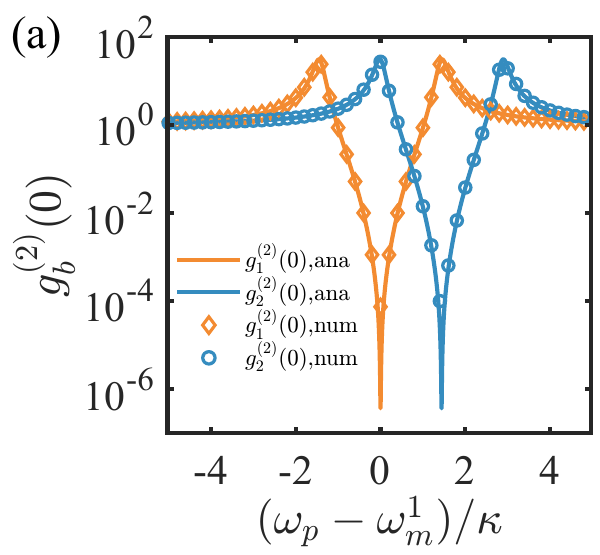}\label{Fig2a}}\hspace{0mm}
   	\centering
   	\subfloat[]{\includegraphics[width=0.47\linewidth]{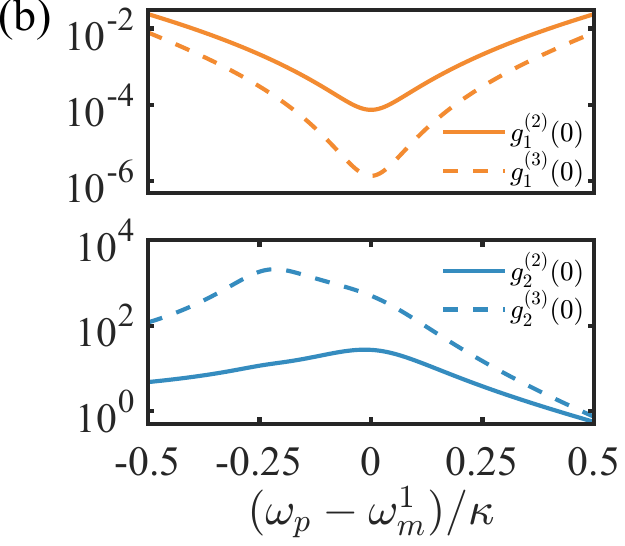}\label{Fig2b}}\hspace{0mm}	
   	\\
   	\vspace{-7mm}  
   	\subfloat[]{\includegraphics[width=0.92\linewidth]{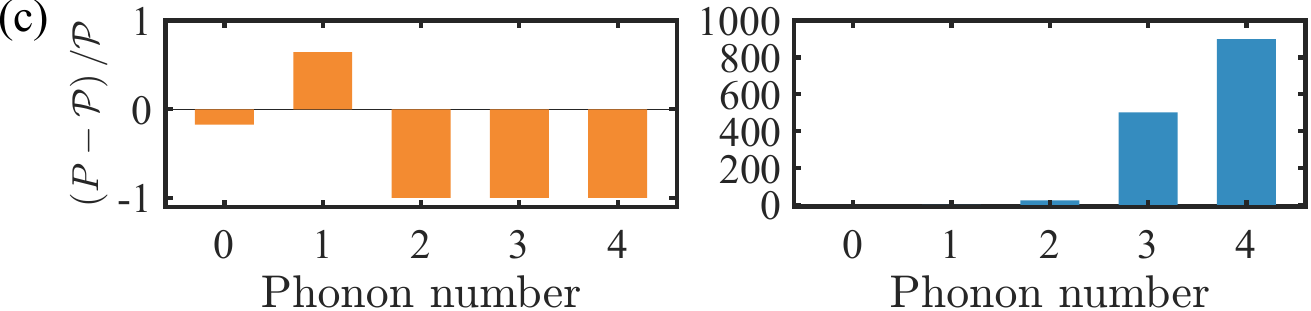}\label{Fig2c}}\hspace{0mm}	
   	\caption{\justifying{(a) Phonon second-order correlation function versus mechanical driving $\omega_p$ for inputs from port 1 [$g^{(2)}_1(0)$] and port 2 [$g^{(2)}_2(0)$]. Solid curves (symbols) correspond to analytical (numerical) results.  Parameters are set to the derived optimal values, with $n_{\mathrm{th}}=0$, $\gamma=\kappa/100$ and $\varepsilon=\kappa/20$. Here, $\omega_m^1$ represents the effective mechanical frequency under the optimal parameters. (b) Second- and third-order correlation functions around the frequency $\omega_p=\omega_m^1$. (c) Deviation of the phonon distribution from a standard Poisson distribution with the same mean phonon number. }}\label{Figure2}
   \end{figure}

   Figure~\ref{Figure2} illustrates the strong nonreciprocal phonon statistics achieved under the optimal system parameters. In Fig.~\subref*{Fig2a}, the second-order correlation functions for pumping through port 1 $(g_1^{(2)}(0))$ and port 2 $(g_2^{(2)}(0))$ are plotted as a function of the mechanical drive frequency. When the mechanical drive is resonant with the effective mechanical frequency for pumping port 1 (i.e., $\Delta_p^1 = 0$), a pronounced dip appears in the correlation function, with $g_1^{(2)}(0) \sim7 \times 10^{-5}$, indicating strong single-phonon blockade. At this same drive frequency, the correlation function for port 2 exhibits a large peak, with $g_2^{(2)}(0) \sim 27$, demonstrating strong phonon bunching. This exceptionally strong quantum nonreciprocity, demonstrated by a 55.8 dB isolation contrast of $g_b^{(2)}(0)$ observed for opposite pump directions, is fundamentally distinct from the nonreciprocity in classical signal transmission rates. Under the condition $\Delta_c \gg |\Delta_F|$, the effective nonlinear couplings are nearly equal ($G_1 \approx G_2$), resulting in almost identical dressed-state energy splittings for both pump directions. This symmetry gives rise to a shift-invariant relationship between the second-order correlation functions, where the curve for $g_2^{(2)}(0)$ is simply a frequency-shifted version of the $g_1^{(2)}(0)$ curve, as observed in Fig.~\subref*{Fig2a}. Consequently, the nonreciprocal phonon statistics can be inverted  by tuning the drive frequency. Tuning the drive to $\omega_p = \omega_m^1 - \sqrt{2}G_2/2$ yields strong phonon blockade for port 2 while simultaneously inducing phonon bunching for port 1, thereby demonstrating flexible, frequency-based control over the directionality of quantum statistics. Fig.~\subref*{Fig2a} demonstrates strong agreement between our analytical model and the numerical results. The minor deviation at the blockade minimum is due to multi-phonon excitations with $m\ge3$ and the neglect of quantum jumps in the wave function approximation method.
   
   To provide a more detailed characterization of the nonreciprocal phonon statistics, we perform further numerical analyses. In Fig.~\subref*{Fig2b} and Fig.~\subref*{Fig2c}, we respectively calculate the third-order correlation function, $g_b^{(3)}(0)=\langle b^{\dagger3} b^3\rangle/\langle b^\dagger b\rangle^3$, and the relative deviation of the phonon number distribution, $\left ( P(m)-\mathcal{P} (m) \right ) /\mathcal{P}(m)$. Here, $P(m)$ is the probability of finding $m$ phonons, and $\mathcal{P} (m)=\frac{\langle b^\dagger b\rangle^m}{m!}\mathrm{exp} \left (  -\langle b^\dagger b\rangle\right )  $ is the standard Poisson distribution. Under the single-phonon resonance condition for port 1 ($\Delta_p^1 = 0$), a pronounced suppression of multi-phonon excitations is observed. The significant reduction in both the second-order ($g_1^{(2)}(0) \approx 7 \times 10^{-5}$) and third-order ($g_1^{(3)}(0) \approx 1.4 \times 10^{-6}$) correlation functions confirms a robust single-phonon blockade.
   This result indicates that the proposed system is a promising candidate for a high-quality single-phonon source when pumped from port 1. According to recent research in photon statistics~\cite{Casalengua2020}, this result naturally arises from the interference between the mechanical coherent component and the squeezed state across all phonon orders. Concurrently, for port 2, we observe $g_2^{(3)}(0) \approx 503 \gg g_2^{(2)}(0) \gg 1$. This behavior characterizes the phonon-induced tunneling phenomenon, which is a purely quantum effect despite exhibiting a classical-like property of super-Poissonian phonon-number statistics. Analogous to the concept of photon-induced tunneling, this phenomenon implies that the absorption of one phonon facilitates the admission of a second or subsequent phonons.  These results are further confirmed by Fig.~\subref*{Fig2c}. When the system is pumped from port 1, we observe $P(1) > \mathcal{P}(1)$ and $P(n) \ll \mathcal{P}(n)$ for $n\ge2$, which are clear signatures of single-phonon blockade. In contrast, when pumped from port 2, the enhanced probability $P(n \ge 2)$ is a distinct indication of phonon-induced tunneling. Therefore, the nonreciprocity in phonon statistics manifests as a stark contrast between two distinct quantum phenomena: strong conventional phonon blockade in one direction of pump light propagation and phonon-induced tunneling in the opposite direction. 

    \begin{figure}[tbp]
   	\centering
   	\subfloat[]{\includegraphics[width=0.9\linewidth]{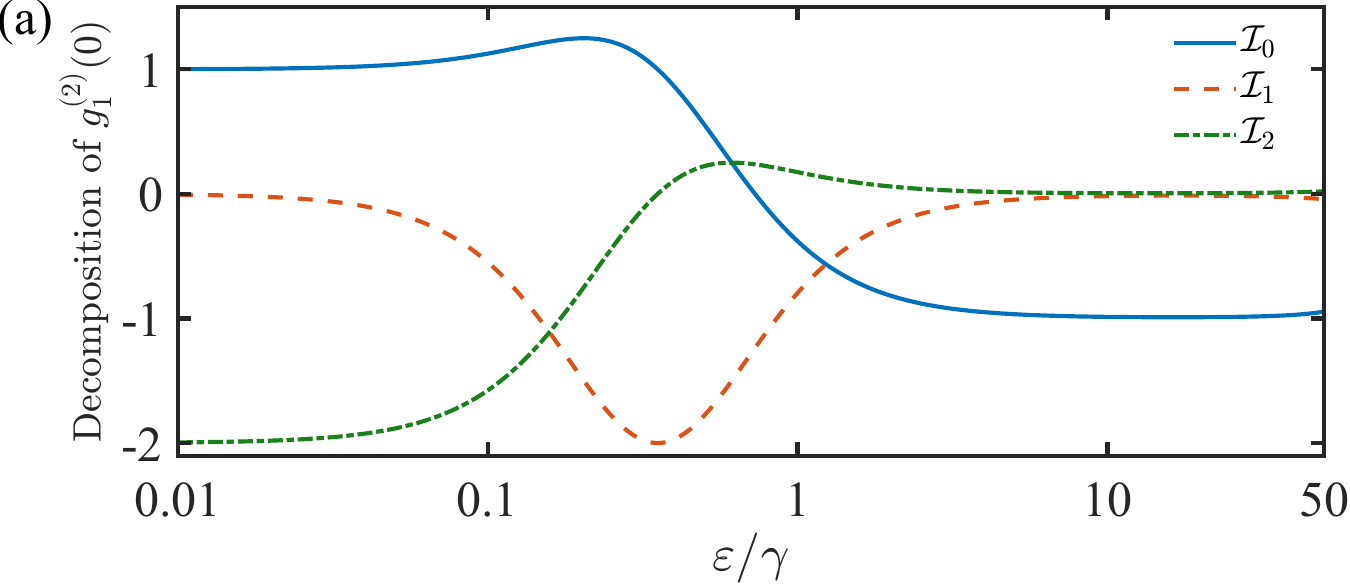}\label{Fig3a}}\hspace{0mm}
   \\
   \vspace{-7mm}
      	\centering
   \subfloat[]{\includegraphics[width=0.9\linewidth]{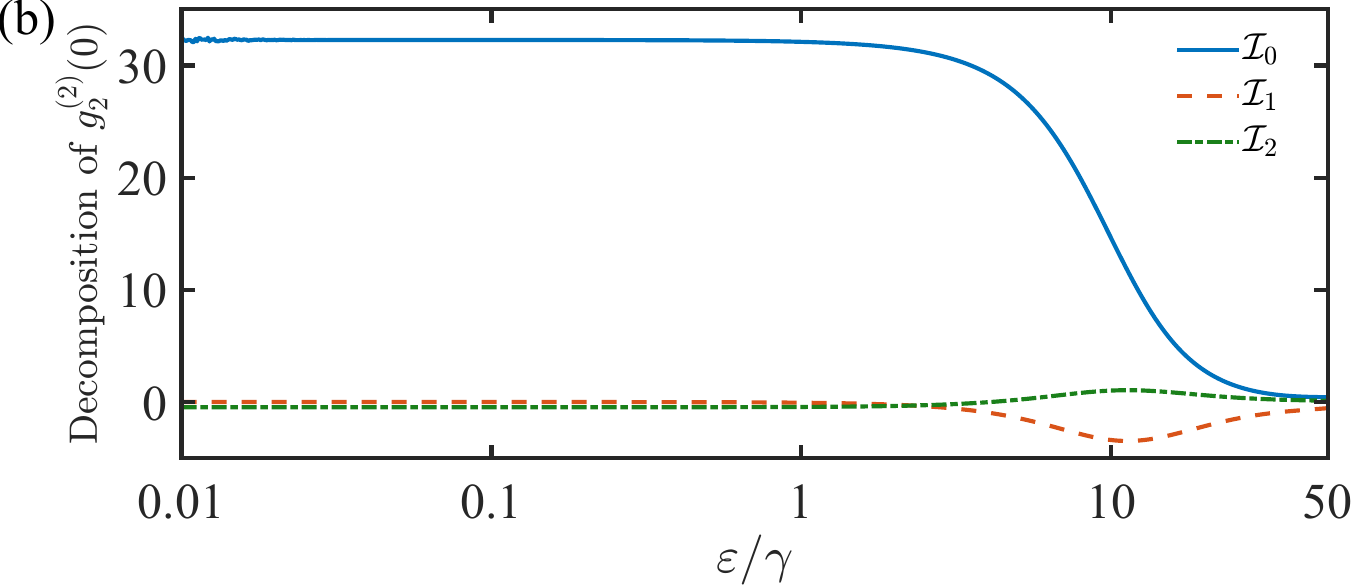}\label{Fig3b}}\hspace{0mm}
   \\
   \vspace{-7mm}
   \centering
     \subfloat[]{\includegraphics[width=0.45\linewidth]{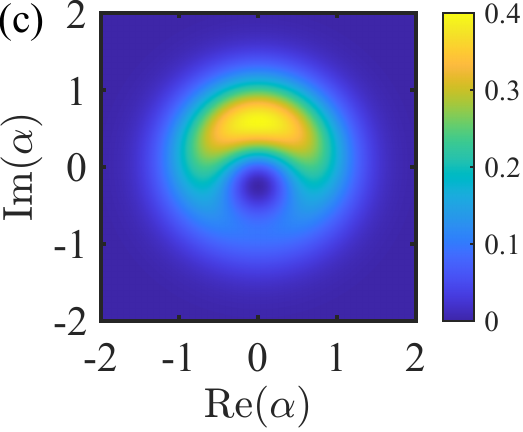}\label{Fig3c}}\hspace{0mm}
        \centering
     \subfloat[]{\includegraphics[width=0.45\linewidth]{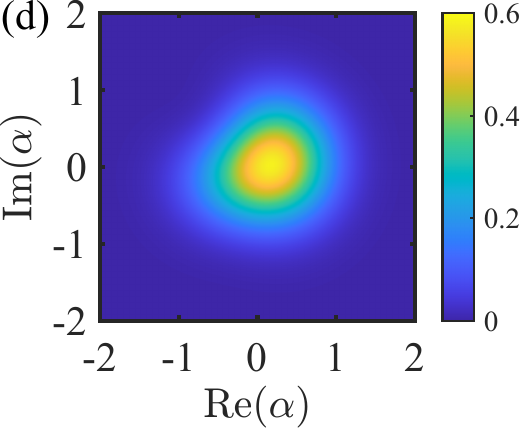}\label{Fig3d}}\hspace{-4mm}
   	\caption{\justifying{(a), (b) Decomposition of the second-order correlation functions $g^{(2)}_1(0)$ [in (a)] and $g^{(2)}_2(0)$ [in (b)] as a function of mechanical driving strength $\varepsilon$ for $\omega_p=\omega_m^1$.  (c), (d) Wigner representations $W_b(\text{Re}(\alpha), \text{Im}(\alpha))$ corresponding to input from port 1 with $\varepsilon/\gamma=1$ [(c)] and input from port 2 with $\varepsilon/\gamma=20$ [(d)]. }}\label{Figure3}
   \end{figure}
   Casalengua et al.~\cite{Casalengua2020} developed a unified framework in which the interference between squeezed quantum fluctuations and a coherent mean field explains both conventional and unconventional photon statistics. Employing this framework, we now analyze the nonreciprocal phonon statistics in our proposed system. By decomposing the phonon operator $b$ into a sum of a coherent component $\langle b\rangle=\beta$ and a fluctuation term $s=b-\beta$, $g_b^{(2)}(0)$ in Eq. (10) can be rewritten as
   \begin{align}
   	g_b^{(2)}(0)=1+\mathcal{I}_0+\mathcal{I}_1+\mathcal{I}_2,
   \end{align}
   according to the decomposition of Ref.~\cite{Casalengua2020}. The term 1 arises from the coherent field, and the incoherent terms are given by 
   \begin{align}
   	&\mathcal{I} _0=\dfrac{\langle s^{\dagger 2}s^2\rangle-\langle s^\dagger s\rangle^2}{\langle b^\dagger b \rangle^2},\\
   	&\mathcal{I} _1=\dfrac{4\mathrm{Re}\left [  \beta^*\langle s^\dagger s^2\rangle\right ]  }{\langle b^\dagger b\rangle^2},\\
   	&\mathcal{I} _2=\dfrac{\langle: X_{s,\phi}^2:\rangle-\langle X_{s,\phi }\rangle^2}{\langle b^\dagger b\rangle^2},
   \end{align}
   where $X_{s,\phi} \equiv (e^{i\phi}s^\dagger +e^{-i\phi}s)/2$ is the quadrature of the fluctuation field $s$ and $\phi \equiv \mathrm{arg}(\beta)$ is the phase of the coherent field. The notation $\mathopen{:}\hat{O}\mathclose{:}$ denotes normal ordering, where all creation operators are placed to the left of annihilation operators. $\mathcal{I}_0>0$ ($\mathcal{I}_0<0$) indicates that the fluctuation field 
   $s$ exhibits super-Poissonian (sub-Poissonian) statistics, $\mathcal{I}_1$ refers to anomalous moments, and a negative $\mathcal{I}_2$ signifies the squeezing of fluctuations. Regarding the phonon antibunching observed with port 1 pumping, Fig.~\subref*{Fig3a} demonstrates that the behavior of these contributions under different driving regimes closely resembles that of resonance fluorescence~\cite{PhysRevLett.125.170402}.
   In the regime $\varepsilon \ll \gamma$, the fluctuation field exhibits its maximal super-Poissonian character with $\mathcal{I}_0 \sim 1$, while simultaneously displaying squeezing with $\mathcal{I}_2 \sim -2$. Consequently, the coherent contribution $1$ is compensated by $\mathcal{I}_0 + \mathcal{I}_2$. The resulting strong antibunching arises from the destructive interference between the coherent state $\beta$ and the super-Poissonian squeezed fluctuations, which occurs to all orders in the phonon numbers. In the regime $\varepsilon \sim \kappa \gg \gamma$, the sub-Poissonian statistics of fluctuation field $s$ are dominated by $\mathcal{I}_0 \approx -1$, while both $\mathcal{I}_1$ and $\mathcal{I}_2$ essentially vanish. Due to the vanishing coherent component (i.e., $\beta \to 0$), the fluctuation field $s$ effectively becomes the total field, thereby determining the overall sub-Poissonian phonon statistics. During the transition between these two driving regimes, analogous to the case of resonance fluorescence~\cite{PhysRevLett.125.170402}, strong antibunching is preserved by the development of skewness in the squeezed fluctuations via the anomalous moment $\mathcal{I}_1$. In this process, a negative $\mathcal{I}_1$ appears as $\mathcal{I}_2\to 0$ and the quantum state evolves from a displaced squeezed thermal state, which conforms to a Gaussian description, to a state with fully non-Gaussian property dominated by the Fock state. To confirm the presence of squeezing in the interference, Fig.~\subref*{Fig3c} displays the Wigner function for $\varepsilon \sim \gamma$. In contrast to the limit $\varepsilon \ll \gamma$ (not plotted due to vanishingly small features), where the small displacement and squeezing are indistinguishable from the dominant thermal background, the Wigner function now reveals a substantial displacement and skewed squeezing. 
     
     We now turn to the decomposition of $g^{(2)}_2(0)$ shown in Fig.~\subref*{Fig3b} for the strong bunching observed when port 2 is pumped. For relatively weak $\varepsilon$, the dominance of the fluctuation term $\mathcal{I}_0\approx 32$ overwhelms the coherent part and drives the system into a super-Poissonian state. With increasing $\varepsilon$, although the super-Poissonian statistics are weakened by the increased total field intensity and the squeezing of fluctuations emerges, the interference between the squeezed fluctuations and the coherent field remains insufficient to reverse these super-Poissonian fluctuations. Thus, the system remains in a phonon bunching regime.  As shown in Fig.~\subref*{Fig3d}, the Wigner function for $\varepsilon/\kappa=1/5$ clearly exhibits both displacement and squeezing.
      \begin{figure}[bp]
   	\centering
   	\subfloat[]{\includegraphics[width=0.85\linewidth]{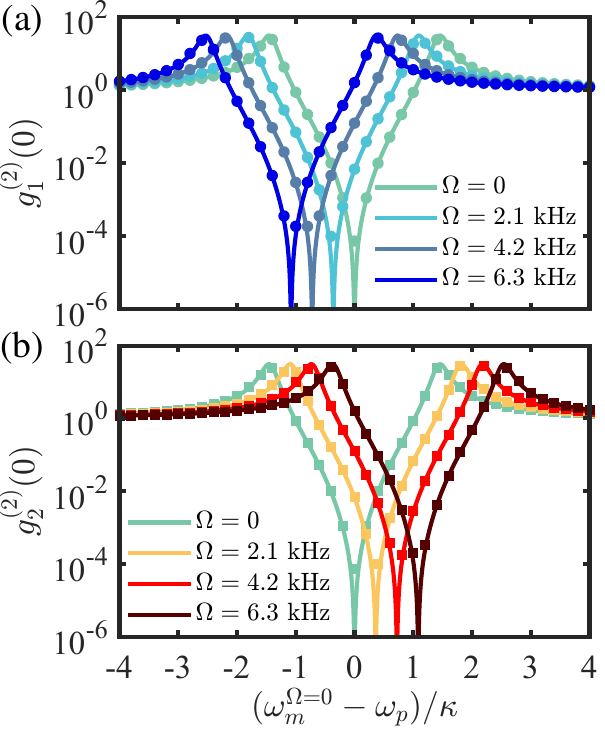}}\\ \vspace{-5mm}%

   	\caption{\justifying{Phonon second-order correlation functions $g^{(2)}_1(0)$ (port 1 input) and $g^{(2)}_2(0)$ (port 2 input) for different spinning angular velocities $\Omega$. Other parameters are the same as in Fig.~\ref{Figure2}. $\omega_m^{\Omega=0}$ denotes the effective mechanical frequency when the resonator is static.}}\label{fig4}
   \end{figure}
  Having elucidated the quantum interference mechanism, we now proceed to investigate how the nonreciprocity depends on the key system parameters, specifically the Sagnac shift. To clearly reveal the direct connection between nonreciprocal phonon statistics and the Sagnac effect, we present $g_i^{(2)}(0)$ at various spin angular velocities $\Omega$, as shown in Fig.~\ref{fig4}. With increasing $\Omega$, a larger Sagnac shift leads to an increase in the effective mechanical frequency $\omega_m^1$, while $\omega_m^2$ decreases, as given by Eqs.~(\ref{eff_mech1_fre}) and~(\ref{eff_mech2_fre}). This results in a blue shift of the single-phonon resonance dip in $g^{(2)}_1(0)$ [see Fig.~\ref{fig4}(a)] and a red shift of the two-phonon resonance peak in $g^{(2)}_2(0)$ [see Fig.~\ref{fig4}(b)]. These two features, PB and PIT, gradually approach each other and eventually coincide at the optimal angular velocity of $\Omega =4.2$ kHz, marking the point of significant nonreciprocity.

   \subsection{Robustness and thermal effects}\label{sec3_c}
\begin{figure}[tbp]
	\centering
	\subfloat[]{\includegraphics[width=0.9\linewidth]{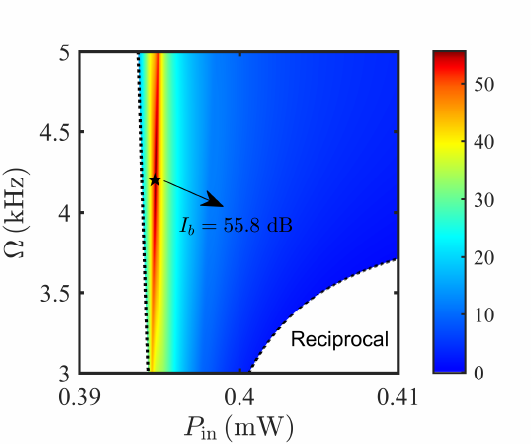}}\\ \vspace{-5mm}%

	\caption{\justifying{Nonreciprocal isolation $I_b$ as a function of pump power $P_{\mathrm{in}}$ and resonator angular velocity $\Omega$. The remaining parameters are consistent with Fig.~\ref{Figure2}. The blank area indicates the reciprocal parameter region.}}\label{fig5}
\end{figure}

   For the practical implementation of nonreciprocal phonon devices, it is crucial to assess the performance of system under realistic experimental conditions. As shown in Fig.~\ref{fig5}, we now analyze the sensitivity of the nonreciprocal isolation $I_b$ to variations in the primary external control parameters, the angular velocity of the cavity rotation $\Omega$ and the pump power $P_{\mathrm{in}}$. We disregard the parameter regions marked as 'reciprocal,' where both pump inputs yield identical phonon statistics (either antibunching or bunching), even if their specific $g^{(2)}(0)$ values differ quantitatively. Our analysis focuses exclusively on the conditions for strong, qualitative nonreciprocity (blockade vs. bunching). As predicted by our theoretical model in Eq.~(\ref{opt_condition}), the most significant nonreciprocity occurs at the optimal parameter set ($\Omega^{\mathrm{opt}}$, $P_{\mathrm{in}}^{\mathrm{opt}}$), which is indicated in Fig.~\ref{fig5} a black asterisk. Importantly, the nonreciprocal isolation $I_b$ exhibits strong robustness against deviations in the angular velocity $\Omega$ from its optimal value  $\Omega_{\mathrm{opt}}$. Quantitatively, an isolation $I_b$ exceeding 50 dB is achieved for angular velocities in the range of $ 3.76 \text{ kHz}  \le \Omega \le  4.76 \text{ kHz}$. Conversely, the pump power $P_{\mathrm{in}}$ requires precise control to maintain significant nonreciprocity. This sensitivity arises because the effective mechanical frequency is strongly dependent on $P_{\mathrm{in}}$ via the optical spring effect. A slight deviation from the optimal power $P_{\mathrm{in}}^{\mathrm{opt}}$ will therefore shift the effective energy levels and render the interaction between states $|n,m\rangle$ and  $|n-1,m+2\rangle$ off-resonant. This combination of features, robustness to $\Omega$ fluctuations and sensitivity to $P_{\mathrm{in}}$, makes our proposed scheme experimentally feasible. In standard experimental setups, achieving precise stabilization of the angular velocity $\Omega$ of a spinning cavity against mechanical jitter is a significant challenge~\cite{Foreman2015,Maayani2018}. Our scheme is inherently robust to this parameter. In contrast, the high sensitivity to pump power $P_{\mathrm{in}}$ can be managed with established technology. High-precision laser power stabilization is routinely achieved using classical feedback loops with acousto-optic modulators (AOMs)~\cite{Kim2007,Balakshy2014,Wang2020,Liu:22}. Furthermore, state-of-the-art nonclassical methods, such as injecting a squeezed vacuum field~\cite{PhysRevLett.121.173601}, have demonstrated laser power stabilization even beyond the standard shot-noise limit.
      \begin{figure}[tbp]
	\centering
	\subfloat[]{\includegraphics[width=0.46\linewidth]{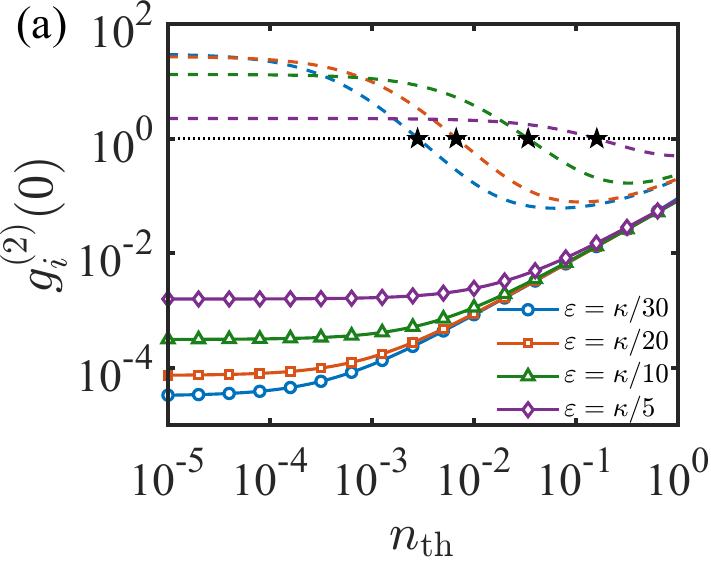}\label{Fig6a}}\hspace{0mm}
	\centering
	\subfloat[]{\includegraphics[width=0.44\linewidth]{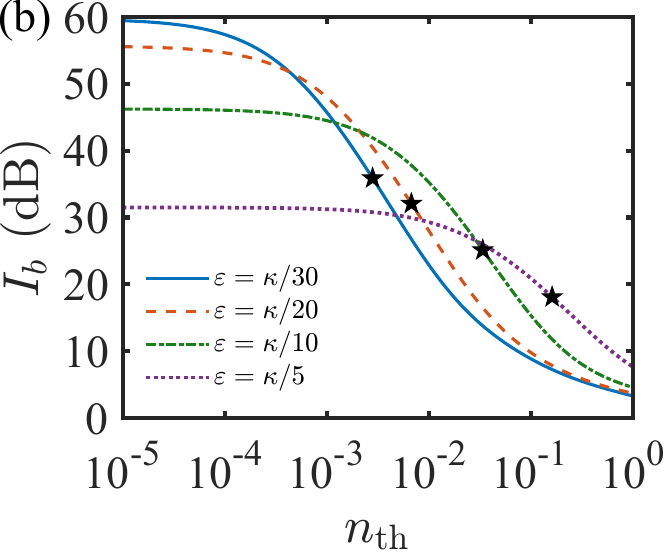}\label{Fig6b}}\hspace{0mm}	
	\\
	\vspace{-7mm}  
	\subfloat[]{\includegraphics[width=0.45\linewidth]{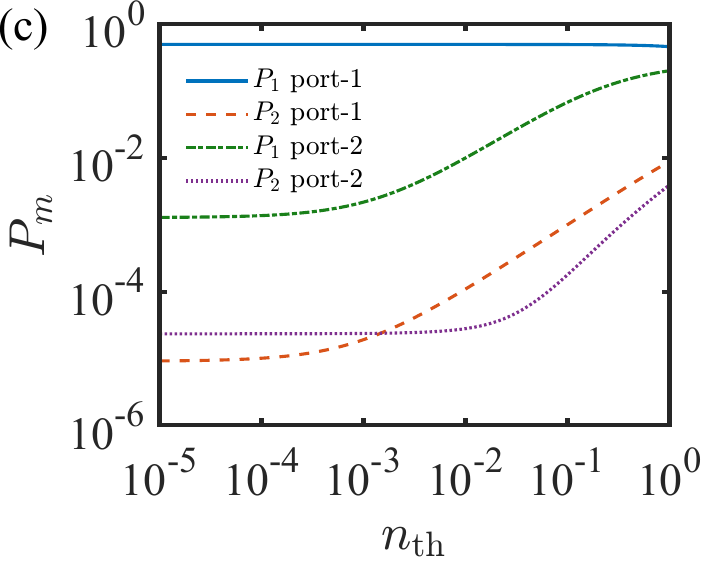}\label{Fig6c}}\hspace{-1mm}	
	\centering
	\subfloat[]{\includegraphics[width=0.45\linewidth]{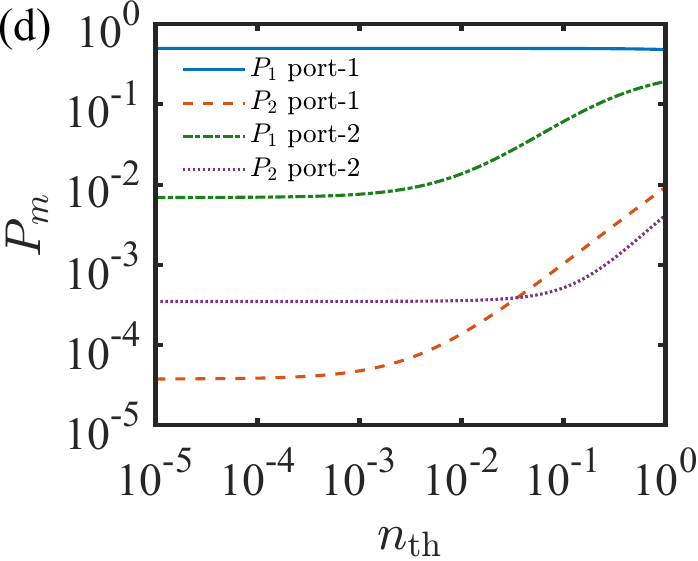}\label{Fig6d}}
	\\
	\vspace{-7mm}
		\centering
	\subfloat[]{\includegraphics[width=0.9\linewidth]{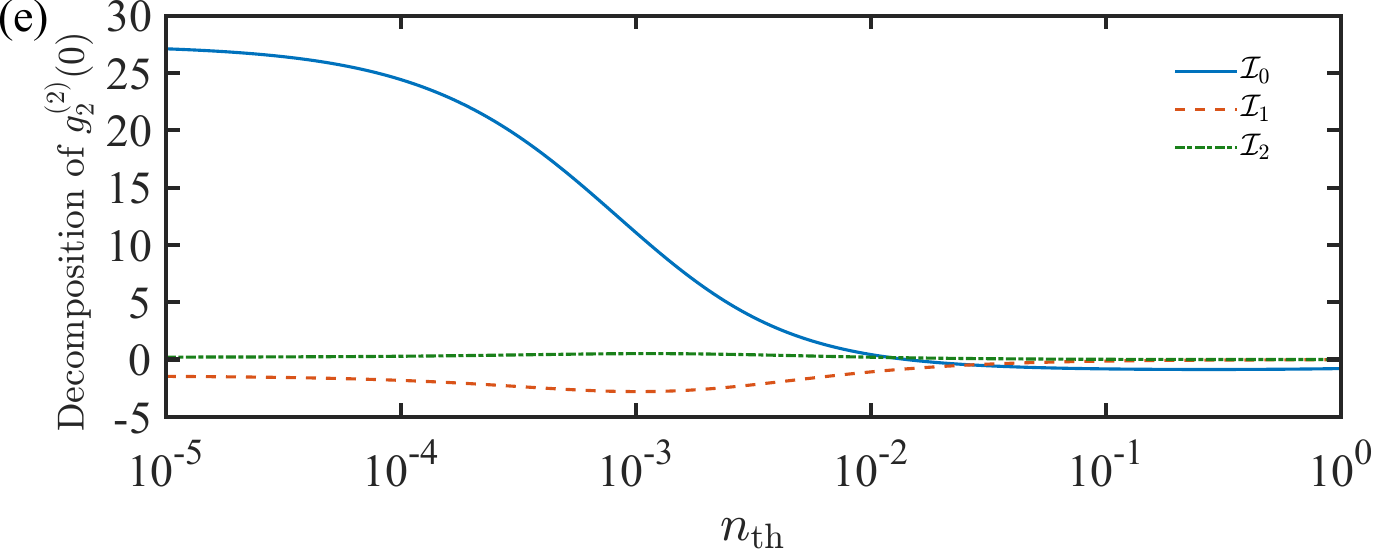}\label{Fig6e}}\hspace{0mm}	
	\caption{\justifying{(a) Second-order correlation functions $g^{(2)}_1(0)$ (solid line) and $g^{(2)}_2(0)$ (dashed line) versus thermal phonon occupation number $n_{\mathrm{th}}$. (b) Nonreciprocal isolation $I_b$ versus $n_{\mathrm{th}}$. The critical points where nonreciprocity vanishes are marked by asterisks. (c), (d) One-phonon ($P_1$) and two-phonon ($P_2$) populations under driving from port 1 and port 2 as functions of $n_{\mathrm{th}}$, with $\varepsilon=\kappa/20$ in (c) and $\varepsilon=\kappa/10$ in (d). (e) Decomposition of $g^{(2)}_2(0)$ as a function of $n_{\mathrm{th}}$. Here, parameters are the same as in Fig. 2, with $\omega_p$ set to $\omega_m^1$. }}\label{Figure6}
\end{figure}

Physically, the reciprocal regimes (the blank areas in Fig.~\ref{fig5}) emerge as a consequence of the breakdown of the strict resoance conditions, which are highly sensitive to the pump-power-dependent optical spring effect. In the left region, this is reflected by the decreased intracavity amplitude $\alpha_i$, which reduces the optical spring shift and consequently causes $\omega_m^i$ to increase. For the forward direction, the system is lifted slightly off the optimal single-phonon resonance, raising $g_1^{(2)}(0)$ while still maintaining its sub-Poissonian property. Meanwhile, the increased $\omega_m^2$ breaks the backward two-phonon resonance ($\omega_p=\omega_m^2+\frac{\sqrt{2}}{2}G_2$) and incidentally drives it toward the single-phonon resonance condition. This progression causes the backward statistics to transition into phonon blockade. With both directions exhibiting phonon blockade, the qualitative statistical nonreciprocity vanishes. Turning to the high-power region on the right, an excessive $P_{\text{in}}$ induces a massive optical spring shift that severely violates the photon-two-phonon resonance condition ($\Delta_a^i \simeq 2\omega_m^i$). Due to this massive detuning, the effective quadratic interaction $A^\dagger b^2 + A b^{\dagger 2}$ becomes highly off-resonant and dynamically suppressed. Without this core nonlinear interaction, both the conventional phonon blockade and the phonon-induced tunneling effects fail, and the phonon statistics in both directions identically degenerate to the standard Poissonian distribution ($g_b^{(2)}(0) \to 1$).

		  We now incorporate mechanical thermal noise into our analysis and investigate its impact on the nonreciprocal phonon statistics. Figs.~\subref*{Fig6a} and~\subref*{Fig6b}  numerically illustrate the dependence of the correlation function $g_i^{(2)}(0)$ and the nonreciprocal isolation $I_b$ on the mean thermal phonon number $n_{\mathrm{th}}$  for various mechanical driving field strengths $\varepsilon$. When the thermal noise reaches a certain level, the correlation functions for both pump directions, as well as the nonreciprocal isolation, begin to experience a detrimental effect. While weaker mechanical driving yields superior nonreciprocity in the zero-temperature limit ($n_{\mathrm{th}} \to 0$) due to reduced multi-phonon probabilities, we find that stronger driving strength $\varepsilon$ significantly enhances robustness against the thermal phonons typically present in practice. In Figs.~\subref*{Fig6a} and~\subref*{Fig6b}, the critical points where nonreciprocity vanishes (i.e., where both pump directions exhibit identical phonon statistics) are marked with asterisks. Specifically, for drive ratios $\varepsilon/\kappa = 1/30, 1/20, 1/10$, and $1/5$, the corresponding critical mean thermal phonon numbers $n_{\mathrm{th}}$ are $2.8\times 10^{-3}$, $6.7\times10^{-3}$, $3.4\times10^{-2}$, and $1.6\times10^{-1}$, respectively. This demonstrates that, by appropriately increasing the mechanical driving strength, the critical $n_{\mathrm{th}}$ can be raised by more than an order of magnitude. Experimentally, pre-cooling the proposed $\omega_m = 2\pi \times 40$ MHz mechanical mode to 10 mK via dilution refrigeration yields a thermal occupation of $n_{\text{th}}^0 \approx 4.7$, which is insufficient to observe the predicted nonreciprocal phonon statistics ($n_{\text{th}} \lesssim 0.1$). By employing the advanced intracavity squeezed-light cooling technology proposed by Gan et al.~\cite{Gan2019Intracavity}, the thermal occupation can be further suppressed to $n_{\text{th}} \sim 2n_{\text{th}}^0/Q_m + \sqrt{n_{\text{th}}^0/Q_m} \sim 7 \times 10^{-3}$ (with $Q_m \sim 10^5$). Such a low thermal phonon number is sufficient to resolve the nonreciprocal phonon blockade in our proposed mechanism. Interestingly, under the influence of environmental thermal noise, the phonon statistics for port 2 transition from bunching to antibunching. In contrast, for port 1, the antibunching is monotonically driven toward the standard Poisson distribution as $n_{\mathrm{th}}$ increases. This behavior reveals an extended  nonreciprocal thermal effect underlying the phonon statistics in our proposed scheme. 
		  
To gain insight into these features, we study the thermal effect in terms of the phonon occupation and the decomposition of $g_b^{(2)}(0)$. As shown in Figs.~\subref*{Fig6c} and~\subref*{Fig6d} , for pump injection through port 1, the single-phonon probability $P_1$ remains almost unaffected by the mean thermal phonon number $n_{\mathrm{th}}$ when $n_{\mathrm{th}} < 1$. In contrast, the two-phonon probability $P_2$ increases with $n_{\mathrm{th}}$ due to enhanced thermal excitation $|1\rangle \to |2\rangle$. As expected, the phonon blockade for port 1 deteriorates with increasing $n_{\mathrm{th}}$. For port 2, although $P_1$ is suppressed by the single-phonon detuning at zero temperature, it is enhanced as $n_{\mathrm{th}}$ rises (within the range $n_{\mathrm{th}}\le 1$). However, the subsequent transition $|0,1\rangle \to |2_\pm\rangle$  requires a sufficiently high-level thermal phonon number to overcome the detuning and become effective. Consequently, the increase in $P_2$ lags behind that of $P_1$, as illustrated in Fig.~\subref*{Fig6c}. This causes $g^{(2)}_2(0) \approx 2P_2/P_1^2$ to undergo a transition from super-Poissonian to sub-Poissonian statistics. Furthermore, by comparing Figs.~\subref*{Fig6c} and~\subref*{Fig6d}, we find that the improved thermal robustness of nonreciprocity at stronger driving $\varepsilon$ arises from the increased stability of the phonon populations $P_1$ and $P_2$ against thermal noise.

To explain the counterintuitive thermal-induced transition from bunching to antibunching observed in Port 2, Fig.~\subref*{Fig6e} illustrates the decomposition of $g^{(2)}_2(0)$ as functions of $n_{\mathrm{th}}$. For the mechanical driving ratio $\varepsilon/\kappa=1/20$, the minimum of $g^{(2)}_2(0)$ occurs at the thermal phonon level $n_{\mathrm{th}}=0.1$, where the components are $\mathcal{I}_0 \approx -0.84$, $\mathcal{I}_1 \approx -0.12$, and $\mathcal{I}_2 \approx 0.02$. While strong quantum fluctuations with super-Poissonian statistics ($\mathcal{I}_0 \approx 27$) dominate the statistics at zero temperature, thermal phonons wash out the super-Poissonian fluctuations and reverse it to the sub-Poissonian regime. The decomposition term $\mathcal{I}_1 < 0$ signifies the presence of a squeezed-coherent component in the state. The overall sub-Poissonian statistics at $n_{\mathrm{th}}=0.1$ results from the coherent contribution 1 being compensated by the sub-Poissonian squeezed fluctuations and the interference between these fluctuations and the coherent component of the system. Notably, the output field resulting from the self-homodyne interference of the coherent component with quadrature-squeezed fluctuations corresponds to a displaced-squeezed thermal state~\cite{Casalengua2020}. This formal equivalence suggests an underlying link to the observed thermal phonon-induced antibunching and warrants a more detailed and precise discussion.

Finally, we examine the viability of the proposed nonreciprocal mechanism under realistic experimental parameters. To concretize this point, we specifically consider an unenhanced bare coupling rate of $g_0/2\pi = 200 \text{ kHz}$, which falls within the typical hundred-kHz regime attainable in practice. Accordingly, to accommodate this weaker coupling strength, the system parameters should be adjusted to $P_{\text{in}} \approx 39.5 \text{ mW}$ and $\Omega \approx 0.84 \text{ kHz}$ according to Eq.~(\ref{opt_condition}). Under these conditions, our calculations show that the nonreciprocal isolation contrast $I_b$ reaches approximately $30 \text{ dB}$. Although the contrast decreases from the idealized $55 \text{ dB}$ limit due to the weaker bare coupling, a $30 \text{ dB}$ isolation still represents a three-order-of-magnitude difference in phonon statistics. This validates that the nonreciprocal transition between PB and PIT remains highly resolvable. Our scheme is experimentally viable with standard traveling-wave WGM resonators, and further performance improvements could be achieved by incorporating advanced phononic-crystal engineering.

	\section{DISCUSSION AND CONCLUSION}\label{section4}
	While the investigation of nonreciprocity in QOM systems was pioneered by Xu et al.~\cite{Xu:20}, our work distinguishes itself by extending the scope of nonreciprocity to the statistical properties of phonons within a spinning QOM system. Specifically, our scheme exhibits significant differences in the following three aspects: (i) Regarding the mechanism, the resonators in Ref.~\cite{Xu:20} are static, and the nonreciprocity arises from directional nonlinear interactions. By optically pumping the system from one side, the effective optomechanical coupling is  coherently enhanced in that direction while remaining inherently weak in the other, thereby inducing nonreciprocity. In our system, nonreciprocity emerges from the interplay between Fizeau drag and the optical spring effect. Here, Fizeau drag first induces asymmetric intracavity intensities, which the optical spring effect subsequently converts into direction-dependent effective mechanical frequencies. (ii) Turning to the nature of the nonreciprocity involved, Ref.~\cite{Xu:20} reports a nonreciprocal photonic platform in which photon statistics and direction-dependent transmission are controlled by the propagation direction of a weak probe field. Our objective is to realize a nonreciprocal phonon source whose statistical properties can be switched by the direction of the strong pump field. (iii) With respect to the nonreciprocal effect, Ref.~\cite{Xu:20} realizes nonreciprocal switching that is limited to transitions between photon blockade (or bunching) and classical statistical regimes. By contrast, our spinning scheme exhibits a nonreciprocal transition between PB and PIT, representing a clear contrast between two purely quantum regimes. Furthermore, an additional nonreciprocal thermal effect is revealed.
	
	It is important to note that the configuration of mutually counter-rotating resonators is crucial for enabling the proposed nonreciprocal mechanism. If the rotation direction of one cavity is reversed such that both cavities co-rotate, the pump field will experience opposite Sagnac-Fizeau shifts in the two cavities. This breaks the frequency degeneracy and disrupts the symmetry of the optical supermodes. Consequently, the perfect cancellation of the first-order linear optomechanical coupling can no longer be achieved, and the effective quadratic coupling is correspondingly suppressed. These combined effects ultimately violate the strict resonance conditions required for high-contrast nonreciprocal phonon blockade and phonon-induced tunneling.
	
	In summary, we have investigated nonreciprocal phonon blockade in a platform with quadratic optomechanical coupling, featuring two spinning WGM resonators coupled to a nanomechanical resonator via the evanescent field. The combined action of Sagnac-Fizeau drag and the optical spring effect induces direction-dependent effective mechanical frequencies, which provides the fundamental basis for nonreciprocal phonon blockade. Consequently, phonon blockade emerges from single-phonon resonance excitation under input from one port, while phonon-induced tunneling arises from a two-phonon resonance transition under input from the other. We also attributed the observed nonreciprocal phonon statistics to the distinct quantum interference between the coherent component and the squeezed fluctuations of the system. Interestingly, we have revealed a counterintuitive thermal effect in which increasing thermal noise can drive a transition from phonon bunching to phonon antibunching.
	Due to the direction-dependent shifts of the dressed-state eigenenergies, our system has the potential to enable nonreciprocal photon blockade by replacing the weak mechanical probe with a weak optical probe, offering a promising route to the design of multifunctional nonreciprocal devices. Beyond the present work, richer nonreciprocal phonon statistics such as two-phonon and unconventional PB in spinning QOM systems merit further investigation.
	
   	\begin{acknowledgments}
   	This work was supported by the National Natural Science Foundation of China (NSFC) under Grants No. 12474353.
   \end{acknowledgments}
 
 	\section*{Data availability}
 		The data that support the findings of this article are available from the authors upon reasonable request.
	\appendix
	\section{DERIVATION OF THE QUADRATIC OPTOMECHANICAL COUPLING}\label{Appendix}
	In this appendix, we provide the detailed derivation from the bare Hamiltonian in Eq.~(\ref{Hamiltonian}) to the diagonalized Hamiltonian in Eq.~(\ref{Eq3}). Under the quasistatic assumption $J \gg \omega_m$, the mechanical displacement $q$ varies slowly compared to the optical field dynamics and can thus be treated as a parameter. The Hamiltonian in Eq.~(\ref{Hamiltonian}) can be rewritten in a matrix form for the optical modes as
	\begin{align}\label{eq_A1}
		H =& \begin{pmatrix} a_1^\dagger & a_2^\dagger \end{pmatrix} 
		\begin{pmatrix} 
			\omega_c \pm |\Delta_F| - g_0 q & J \\ 
			J & \omega_c \pm |\Delta_F| + g_0 q 
		\end{pmatrix} 
		\begin{pmatrix} a_1 \\ a_2 \end{pmatrix} \notag\\
		&\quad+ \frac{1}{2}\omega_m(p^2 + q^2).
	\end{align}
	Subsequently, the matrix in the Eq.~(\ref{eq_A1}) can be diagonalized as
	\begin{equation}
		H = \begin{pmatrix} a_+^\dagger & a_-^\dagger \end{pmatrix} 
		\begin{pmatrix} 
			\omega_+(q) & 0 \\ 
			0 & \omega_-(q) 
		\end{pmatrix} 
		\begin{pmatrix} a_+ \\ a_- \end{pmatrix} 
		+ \frac{1}{2}\omega_m(p^2 + q^2).
	\end{equation}
	Here, the two normal modes in this diagonal representation are linear superpositions of the bare modes $a_1$ and $a_2$, which are expressed as
		\begin{align}
			a_+ &= \frac{1}{A_+} \left[ J a_1 + \left( g_0 q + \sqrt{J^2 + (g_0 q)^2} \right) a_2 \right], \\
			a_- &= \frac{1}{A_-} \left[ J a_1 + \left( g_0 q - \sqrt{J^2 + (g_0 q)^2} \right) a_2 \right].
		\end{align}
The coefficients $A_\pm$ are the corresponding normalization factors, which are given by
\begin{equation}
	A_\pm^2 = J^2 + \left( g_0 q \pm \sqrt{J^2 + (g_0 q)^2} \right)^2.
\end{equation}
Due to the tunneling interaction between the two bare modes $a_1$ and $a_2$ and the optomechanical coupling, the normal modes $a_+$ and $a_-$ experience mode splitting and an avoided energy crossing. The corresponding eigenfrequencies are expressed as
\begin{align}
	\omega_{\pm}(q) = \omega_c \pm |\Delta_F| \pm J\sqrt{1 + \left(\frac{g_0 q}{J}\right)^2}.
\end{align}
Here, the sign of the Sagnac frequency shift is determined by the pump input direction and is independent of the normal mode subscripts. In the limit $J \gg g_0 q$, Taylor expansion up to the lowest non-vanishing order of the displacement $q$ gives
\begin{align}
	\omega_{\pm}(q) \approx \omega_c \pm |\Delta_F| \pm \left(J + \frac{g_0^2}{2J}q^2\right).
\end{align}
By defining $G_0 = g_0^2/(2J)$, we recover the effective quadratic optomechanical Hamiltonian in Eq.~(\ref{Eq3}) of the main text.
	\bibliography{Primary_manuscript}

	
	
	

\end{document}